\documentclass[lettersize,journal]{IEEEtran}
\usepackage{amsmath,amsfonts}
\usepackage{algorithmic}
\usepackage{algorithm}
\usepackage{array}
\usepackage[caption=false,font=normalsize,labelfont=sf,textfont=sf]{subfig}
\usepackage{textcomp}
\usepackage{stfloats}
\usepackage{url}
\usepackage{verbatim}
\usepackage{graphicx}
\usepackage{cite}
\usepackage{xcolor}
\begin{document}

\title{Self-Supervised Learning for Enhancing Angular Resolution in Automotive MIMO Radars}

\author{Ignacio Roldan,~\IEEEmembership{Graduate Student Member,~IEEE,}, 
Francesco Fioranelli,~\IEEEmembership{Senior Member,~IEEE,}
Alexander Yarovoy,~\IEEEmembership{Fellow,~IEEE}

\thanks{Copyright (c) 2015 IEEE. Personal use of this material is permitted. However, permission to use this material for any other purposes must be obtained from the IEEE by sending a request to pubs-permissions@ieee.org. 

I. Roldan, F. Fioranelli, and A. Yarovoy are with the Microwave Sensing Signals \& Systems (MS3) Group, Department of Microelectronics, TU Delft, 2628 CD, The Netherlands (email: i.roldanmontero@tudelft.nl; f.fioranelli@tudelft.nl; a.yarovoy@tudelft.nl)}
}
\markboth{ACCEPTED FOR PUBLICATION AT IEEE TRANSACTIONS ON VEHICULAR TECHNOLOGY, APRIL, 2023}%
{Shell \MakeLowercase{\textit{et al.}}: A Sample Article Using IEEEtran.cls for IEEE Journals}


\maketitle

\begin{abstract}
A novel framework to enhance the angular resolution of automotive radars is proposed. An approach to enlarge the antenna aperture using artificial neural networks is developed using a self-supervised learning scheme. Data from a high angular resolution radar, i.e., a radar with a large antenna aperture, is used to train a deep neural network to extrapolate the antenna element's response. Afterward, the trained network is used to enhance the angular resolution of compact, low-cost radars. One million scenarios are simulated in a Monte-Carlo fashion, varying the number of targets, their Radar Cross Section (RCS), and location to evaluate the method's performance. Finally, the method is tested in real automotive data collected outdoors with a commercial radar system. A significant increase in the ability to resolve targets is demonstrated, which can translate to more accurate and faster responses from the planning and decision-making system of the vehicle.
\end{abstract}

\begin{IEEEkeywords}
automotive radar, MIMO, angular resolution, neural networks, machine learning, radar signal processing.
\end{IEEEkeywords}

\section{Introduction}
\IEEEPARstart{R}{adar} systems are used nowadays for many applications of advanced driver assistance systems (ADASs), such as lane-change assist, adaptive cruise control, or forward collision avoidance. Moreover, radar has unique features that make it a key technology for reaching highly automated driving (HAD), a capability most automotive companies are trying to achieve. With a radar sensor it is possible to measure the range, radial velocity, and angle of the targets in the scene. However, these systems do not yet fulfill the automotive industry's requirements \cite{Bilik2019}, especially when looking at the current capabilities regarding angular resolution.

High angular resolution is crucial for fully autonomous driving. Firstly, automotive radars need to have high azimuth discrimination capability to separate targets on the road located at the same distance and moving at the same speed. Secondly, high elevation resolution is needed to discriminate which objects can be driven over, such as small debris on the road or speed bumps, which objects can be driven under, such as bridges or tunnel entrances, and which objects should be safely avoided as significant obstacles on the road for the vehicles. Finally, typical automotive targets have multiple scattering points due to their different curvatures and corners \cite{Buddendick2012, Buhren2006, Andres2012}. A high-angular resolution system makes it possible to capture them, leading to denser point clouds. These richer representations can help for subsequent processing steps in the signal processing pipeline, such as target classification \cite{Palffy2020,Perez2018,Kim2020, Cai2021}, road mapping \cite{Lundquist2011,Orr2021}, or scene semantic segmentation \cite{Schumann2020}.

The use of larger antenna arrays would increase the angular resolution, but it would be unfeasible to integrate them into conventional vehicles due to their size and cost. For this reason, commercial automotive radars use Multiple input Multiple output (MIMO) \cite{stoica_li_2009, Bergin2018, Li2007} radar principles to achieve higher angular resolution without increasing the size of the system. A MIMO radar can synthesize virtual arrays with large apertures with only a few transmit and receive antennas.

The basic principle for angle estimation in MIMO radars relies on the extra distance that a signal reflected from the same target must travel to reach different antennas in the system. The easiest way to exploit this, known as Digital Beam Forming (DBF), applies a Fourier transform to translate the time delay of received signals at each antenna into a phase shift, which is proportional to the angle of arrival of the signal. There are more advanced algorithms such as MVDR \cite{Capon1969}, MIMO-Monopulse \cite{Feng2018}, subspace methods such as MUSIC \cite{Schmidt1986} or ESPRIT \cite{Roy2009}, or methods based on compressive sensing \cite{Candes2014, Fortunati2014, Heckel2016}. However, these methods usually require higher computational costs and longer integration times; hence, they are not always easily applicable in automotive scenarios. Moreover, the angular resolution achieved with all of them is still proportional to the number of virtual antennas of the system \cite{Sun2020}. Therefore, a direct way of improving angular resolution would be to increase the number of transmitters and receivers of the system, which is the trend the automotive radar industry is following \cite{Bilik2018,Brisken2018,Stolz2018}. For example, the prototype presented in \cite{Stolz2018} has 4 transmitting elements and 16 receiving elements, or the Texas Instruments MIMO radar \cite{Tidep} used in this work with 12 transmitters and 16 receivers. However, this increase also raises the price and the size of the radar, both critical constraints in the automotive industry.

A recent approach to address this problem is to exploit the relationship between the responses of each virtual antenna to extrapolate new responses of artificial antennas not physically present in the system. Correctly estimated, these new responses can be used in combination with the aforementioned algorithms, yielding a higher angular resolution. An example of this approach has been studied in \cite{Alistarh2021}, where piecewise cubic extrapolation is used, and in \cite{Cho2015}, by using Autoregressive (AR) models in a Front-Looking SAR geometry. Also, in \cite{Cho2021}, the authors propose to apply a Guided Generative Adversarial Network (GGAN) to range-angle radar matrices to produce a higher resolution version of them. However, this method is used after the angle estimation algorithm; therefore, the data is simply treated as real-valued images. The related results in the state of the art are mostly verified with simulated or point-like targets in very simple scenarios. Moreover, the undesirable effects induced by these extrapolation methods (e.g., the creation of ghost targets or artifacts, or the loss of real targets) are not studied in detail, despite being critical in an automotive scenario.

This paper presents a novel framework to enhance the angular resolution of MIMO radar systems without increasing the physical number of antennas. The proposed framework uses the recent advances in machine learning techniques to train a Neural Network (NN) with data from a large aperture radar (i.e., with high angular resolution) in a self-supervised scheme. Afterward, the trained network can be employed to enhance the angular resolution of a smaller aperture radar in the operational stage, without using the larger radar used to gather training data. The main contributions of this work are:

\begin{itemize}

\item Introduction of a novel framework for enhancing angular resolution in automotive MIMO radars that use self-supervised learning to train a NN in a teacher-student fashion between two radar systems, one used in the design phase and one in operations.
\item Design, train, and evaluate the NN performances using 1 million simulated scenes. This includes the verification of the results in terms of angular accuracy and resolution using different DoA estimation algorithms, as well as the undesirable effects that the extrapolation may induce.
\item Verification of the framework in real outdoor automotive scenarios, where the targets are moving, in contrast with the simulated or ideal scenarios found in the literature proposing alternative methods for increasing angular resolution with extra artificial antennas \cite{Alistarh2021, Cho2015}. 

\end{itemize}

The rest of the paper is organized as follows. Section II presents the relevant background for this work. Section III introduces the proposed self-supervised framework with details of the NN used. Section IV presents results using 1 million simulated scenes and relevant radar metrics. Section V analyses the performances using real data collected with a commercial automotive radar, first with corner reflectors and then with people. Finally, conclusions are drawn in Section VI.

\section{MIMO Radar Background}
The basic principle for angle estimation in a MIMO radar relies on the extra distance that a signal must travel to reach different antennas. In the general case, an array of M antennas receiving signals generated by P sources is considered. The signal observed by the mth antenna is the sum of the time-delayed versions of the original signal:

\begin{equation}
    x_m (t)=\sum_{n=1}^{P}s_p (t-\tau_{mp}) + n_m(t),
    \label{signalModel}
\end{equation}
where:
\begin{itemize}
\item $s_p (t)$ is the signal scattered by the target $p$ received by the first antenna.
\item $\tau_{mp}$ is the propagation time difference between the instant the first antenna received the signal and the instant the $m_{th}$ antenna received the signal.
\item $n_m$ is the additive noise.
\end{itemize}

\begin{figure}
\includegraphics[width=\linewidth]{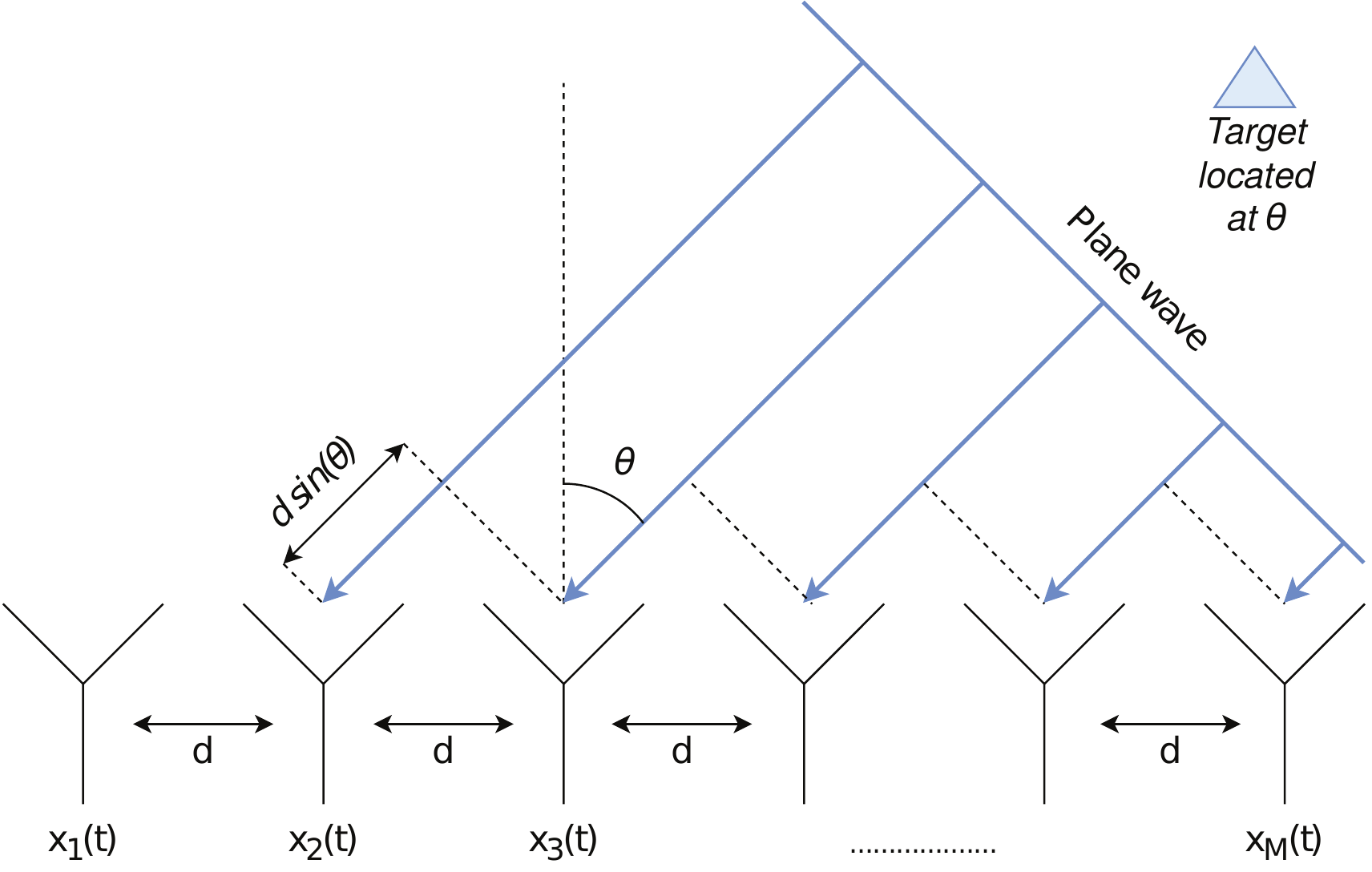}
\caption{A wave scattered from a target located at $\theta$, arriving at a MIMO array. The time delay between each antenna element depends on the target angle $\theta$.}
\label{MIMODiagram}
\end{figure}

The time delay $\tau_{mp}$ is associated with a specific wave p (i.e., the reflection of a single target) and can be translated to a phase shift. In general, these phase shifts induced by the time delay can be exploited to estimate a scattered wave's Direction of Arrival (DoA). A visual representation of how the time delay changes with the DoA can be seen in Fig.~\ref{MIMODiagram}. The time delay between elements for that specific array distribution is $d \sin{\theta} / c$, where $c$ is the speed of light, $d$ is the distance between antennas and $\theta$ is the relative angle or DoA of the target with respect to the radar. Therefore, different values of DoA will lead to different time delays.

For a uniform linear array (ULA), like the one presented in Fig.~\ref{MIMODiagram}, the Rayleigh beamwidth of a conventional beamformer (i.e., the distance between the first two nulls of the beam pattern) is approximately given by \eqref{resolution}  \cite{CHUNG2014599}. Thus, the angular resolution of the system will be inversely related to the number of virtual antennas $M$, which in MIMO systems is provided by the product of the number of transmitting and receiving antennas.

\begin{equation}
    \theta_{BW}  \approx  \frac{\lambda}{Md \cos{\theta}}
    \label{resolution}
\end{equation}

However, it is clear from \eqref{signalModel} that the signals received with each antenna $x_m (t)$ with m=1,..,M are highly correlated. Using this correlation, it is possible to extrapolate a new signal $x_{M+1} (t)$, essentially the signal the M+1 antenna would receive if present in the system. If adequately extrapolated, this additional signal can be used to increase the angular resolution of the system. For this reason, this research proposes a method to increase the aperture of the antenna array without adding physical antennas, by extrapolating the spatial signals to improve the angular resolution of MIMO systems. This method is described in the next section.

\section{Proposed Method}
As explained in the previous section, the signals $x_m (t)$ perceived by each antenna of a MIMO radar receiver are the sum of time-delayed versions of the original signal. Thus, the response of each element is highly correlated with the neighbor elements. However, this relationship can be very complex if many targets at different DoA values are present in the scene. Nevertheless, state-of-the-art automotive radars have high resolution in both range and Doppler, which leads to a very sparse angle-space representation when looking at individual range-Doppler cells as a function of azimuth angle. 

\begin{figure}
\includegraphics[width=\linewidth]{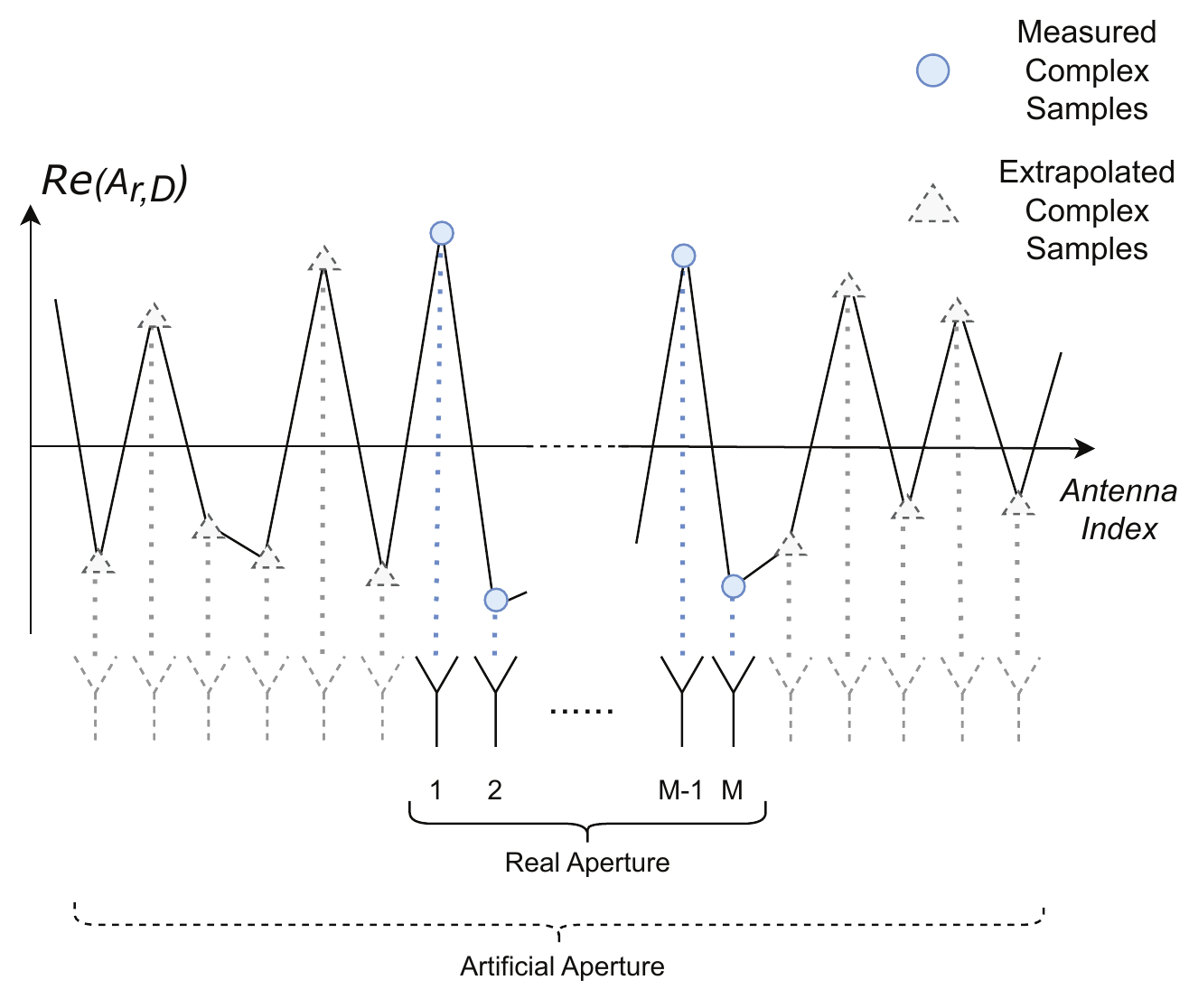}
\caption{Scheme of the extrapolation of the antenna vector for a fixed range-Doppler-time cell. For simplicity, this is shown only for the real component of $A_{r,D}$ but the same procedure is applied to the imaginary part.}
\label{Extrapolation}
\end{figure}

After down-converting, filtering, sampling, and FFT processing, each $x_m (t)$ signal can be arranged in a range-Doppler matrix. These matrices can be stacked to form a radar cube with dimensions range-Doppler-antenna. For a fixed range-Doppler tuple, the resulting antenna vector $A_{r,D}$ can be used to compute an angular profile using any of the angle estimation methods mentioned in section I. However, before estimating the angle, the aforementioned correlation of this vector can be exploited by the proposed method to extrapolate additional complex samples, as illustrated in Fig.~\ref{Extrapolation} for the real-part component.

\begin{figure}
\includegraphics[width=\linewidth]{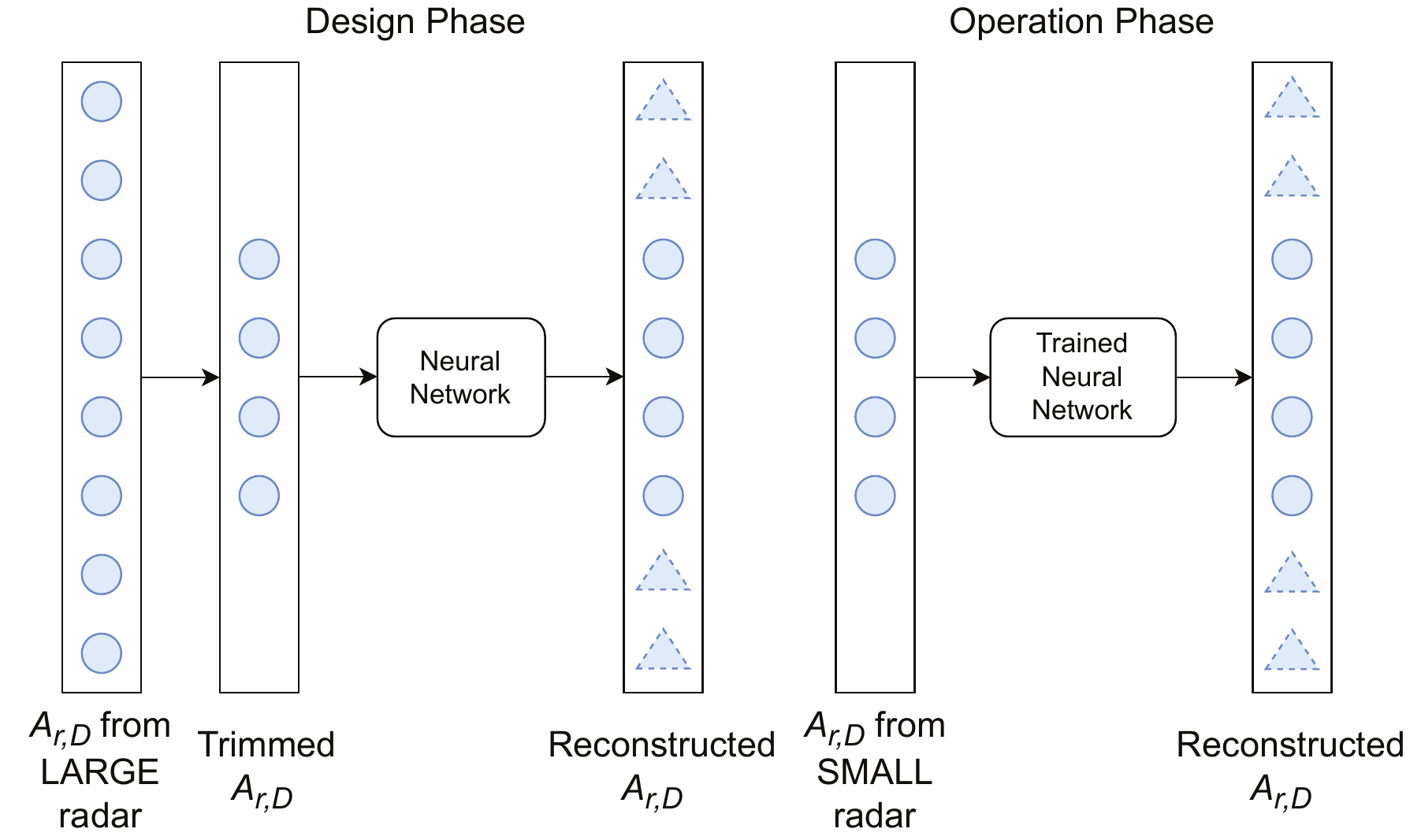}
\caption{A high-level diagram of the proposed approach. In the design phase, data from a high-resolution MIMO radar ('\textit{large}') is used to train a NN that can extrapolate full antenna vectors from subsampled antenna vectors. In the operational phase, the trained NN can be applied directly to data from a low-resolution MIMO radar ('\textit{small}').}
\label{HighLevelDiagram}
\end{figure}

Time series extrapolation has been extensively studied in the literature \cite{brockwell2016introduction}. Many statistical models are available to describe the likely outcome of a time series in the immediate future, such as AR models. However, these statistical models have been recently outperformed by the rise of Neural Networks (NNs) if enough diverse data is available. In this work, a NN is used to enlarge the antenna aperture by extrapolating the time series formed by the response in each antenna element. The proposed approach consists of two phases, namely, the design phase and the operational phase. In the design phase, a high angular resolution radar (e.g., a MIMO radar with extra number of transmit channels which results in a larger aperture of the virtual array in comparison to the operational radar) is used to collect data. Then, this data are used to train a NN proposed in a self-supervised scheme to forecast the response of antenna elements additional with respect to the ones in the operational radar (e.g., a MIMO radar with less number of transmit antennas and, as a result, a smaller aperture of the virtual array). In the operational phase, the NN is used to enhance the angular resolution of an operational radar by forecasting the response of extra antenna elements. It is important to notice that the radar with a large virtual aperture is needed only during the design phase and that radar will not be used in the operational phase. In the operational phase, only the operational radar on board the vehicles will be used, in conjunction with the pre-trained NN to enhance its performance. A visual representation of this approach can be seen in Fig.~\ref{HighLevelDiagram}.

The approach proposed in this work is independent of the radar's waveform, but the algorithm's start point must be a 3D range-Doppler-number of antenna data cube. In most cases in automotive, an FMCW radar system is used, and therefore a 2D FFT can be applied to obtain range-Doppler data \cite{Bilik2019, Dickmann2016}. However, given the sparse nature of the data, most of the cells will contain only noise and must be filtered before applying any extrapolation algorithm. A Cell Averaging Constant False Alarm Rate (CA-CFAR) \cite{Richards2015} detector is applied for this, and only the range-Doppler cells containing at least one target will be used. Finally, the remaining antenna vectors must be subsampled to match the number of virtual antennas of the low-resolution radar envisaged to be used in the operational phase after training.
Then, a NN can be trained using the subsampled antenna vectors as input, and the extreme samples of the antenna vectors as labels (i.e., the samples that must be extrapolated). Therefore, the NN is trained in a self-supervised manner, where the labels are generated automatically from the data. 
In summary, the steps of the proposed pipeline are:
\begin{enumerate}

 \item Apply classical radar signal processing to the data obtained with a large high angular resolution radar to get range-Doppler-antenna data cubes. In this work, a Hamming window and a 2D FFT have been used.
 \item Apply a detector to filter the antenna vectors that only contain noise. In this work, a CA-CFAR detector has been used.
 \item Subsample the resulting antenna vectors, taking the inner L samples corresponding to the number of antenna elements of the low angular resolution radar that will be used in the operational phase.
 \item Generate the labels for the design phase by taking the K outer samples of the antenna vectors, where K+L = M, being M the number of virtual antennas of the high resolution radar.
 \item Train the NN in a self-supervised scheme, using the subsampled antenna vectors as input and the extreme samples of the antenna vectors as labels.
 \item Finally, in the operational/testing phase, data captured with a small aperture radar can be processed with the NN to enlarge its aperture, and thus enhance its angular resolution.
\end{enumerate}

\begin{figure*}[!t]
\centering
\includegraphics[width=\textwidth]{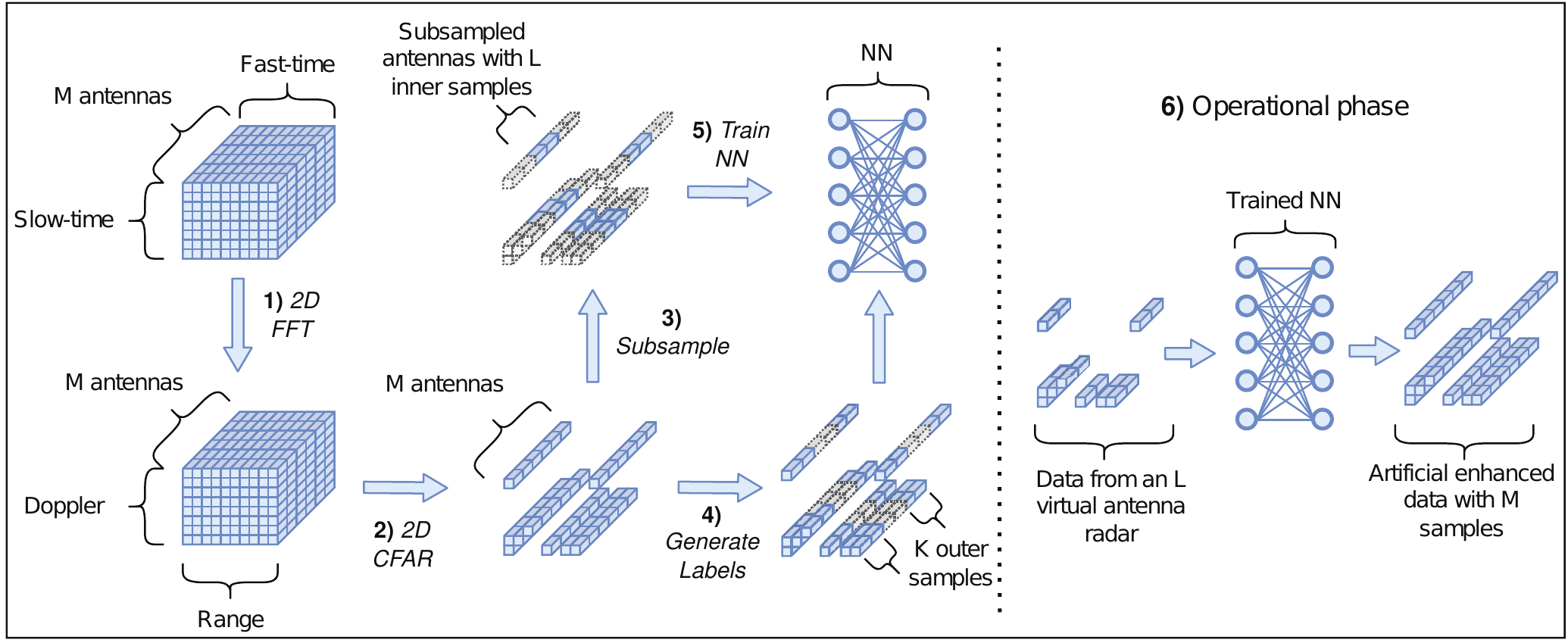}
\hfil
\caption{Block diagram of the proposed method. First, a 2D FFT (1) is applied to the data, followed by a detection stage (2). The remaining antenna vectors (i.e., only those containing targets) are subsampled to generate the training data (3) and the corresponding labels (4). Afterwards, the NN is trained (5). Finally, the trained NN can be used to extrapolate the aperture of a compact radar at the operational phase (6).}
\label{BlockDiagram}
\end{figure*}

A block diagram of the method can be seen in Fig.~\ref{BlockDiagram}. It is important to notice that the proposed method is performed before the DoA estimation algorithm, essentially in order to generate additional artificial antennas from an extrapolation of the available raw radar signals. Thus, any DoA estimation technique, such as MUSIC, MVDR, or a simpler FFT-based beamformer, can be applied after the proposed method. Because of its easy implementation, the simple Fourier beamformer \cite{CHUNG2014599} has been used as DoA estimation technique for most of the results of this work when evaluating the enhanced performance provided by the proposed method. However, the single snapshot MUSIC algorithm presented in \cite{LIAO201633} is also implemented to verify that the proposed approach is estimator agnostic.

\begin{figure}
\includegraphics[width=\linewidth]{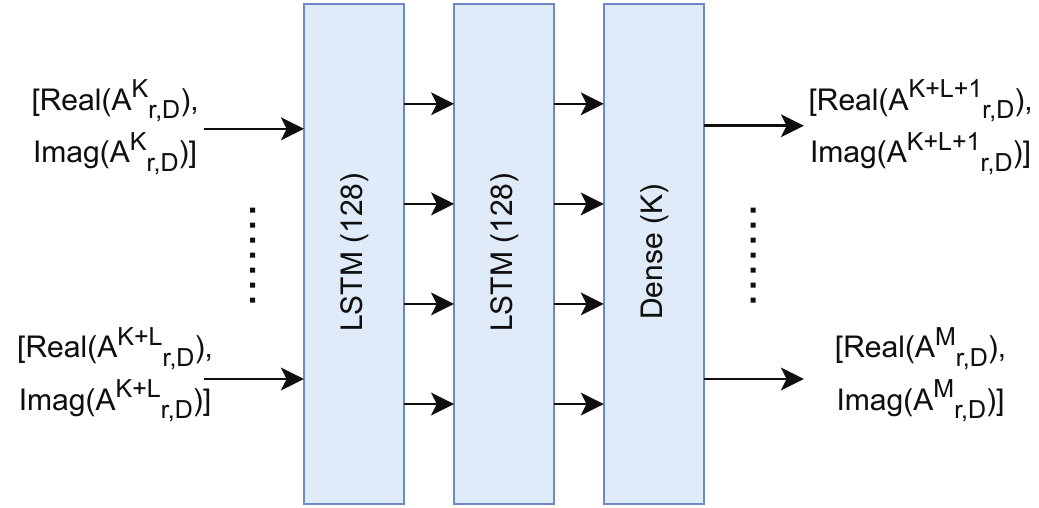}
\caption{Neural Network architecture used with two LSTM layers stacked. The input is an L$\times$2 matrix, while the input is a K$\times$2 matrix.}
\label{LSTM}
\end{figure}

The most common NN model for dealing with sequential data is the so-called Long Short-Term Memory (LSTM) architecture. The main idea is to incorporate a memory cell that can maintain its state over time, capturing long-term dependencies. LSTMs have been proved state-of-the-art in many different domains \cite{Greff_2017}. In this case, each new element will be predicted by conditioning on the joint probability of previous values, including the past predictions. For this reason, the network learns to make conservative predictions and avoid successive errors accumulating rapidly, causing the signal to diverge. The proposed model has two stacked LSTM layers with 128 units each, followed by a dense layer. A block diagram of the network can be seen in Fig.~\ref{LSTM}. Each $A_{r,D}$ has been normalized only in its absolute value within the range of 0 to 1, while at the same time taking care that the phase information relevant for subsequent DoA estimation is maintained. This normalization, a common practice in machine learning, is performed to speed up the convergence of the NN. For the extrapolation of the initial samples $A_{r,D}^i$ with $i \in [1,K]$  the complex conjugated of the 'flipped' signal has been used. Essentially, the flipping operator reverses the order of samples and can be interpreted as the mirrored sequence with respect to the middle element. The real and imaginary parts have been concatenated in a new dimension, treating the problem as a multivariate time series extrapolation.

\section{Evaluation on Simulated Results}

A MIMO radar has been simulated to evaluate the performance of the presented method. The transmitter and receiver arrays have been placed closely located so that targets in the far-field are at the same distance from both arrays. The transmitted waveforms are assumed to be mutually orthogonal, and the individual elements are isotropic antennas for the $y > 0$ half-plane. The resulting virtual array is a Uniform Linear Array (ULA) with $\lambda /2$ separation between elements. One million scenes have been generated in a Monte Carlo fashion, where different number of point targets have been placed in the angular space, assuming they are at the same range bin. The RCS and location of the targets have been sampled from a uniform distribution. Table \ref{SimulationParameters} summarises the parameters of the simulation.

\begin{table}[!t]
\caption{Simulation Parameters\label{SimulationParameters}}
\centering
\begin{tabular}{|c|c|}
\hline
Number of scenes & $10^{6}$\\
\hline
Number of targets per scene & $ X \sim \mathcal{U}(1,10)$\\
\hline
Position of the targets ($\deg$) & $ X \sim \mathcal{U}(-70,70)$\\
\hline
RCS of each target (dB) & $ X \sim \mathcal{U}(0,10)$\\
\hline
SNR in the scene (dB) & $ X \sim (-5, 0, 5, 10, 15, 20, 25)$\\
\hline
M = Num virtual antennas large array & 86\\
\hline
L = Num virtual antennas small array & 44\\
\hline
\end{tabular}
\end{table}

The LSTM network has been trained using 900.000 scenes, leaving 50.000 for validation and 50.000 for testing. Adam optimizer has been employed using the default hyperparameters ($\eta$=0.001, $\beta_1$=0.9, $\beta_2$=0.999, $\epsilon$=1e-7). The results shown in the rest of the section have been computed using only the test data set.

\begin{figure}
\includegraphics[width=\linewidth]{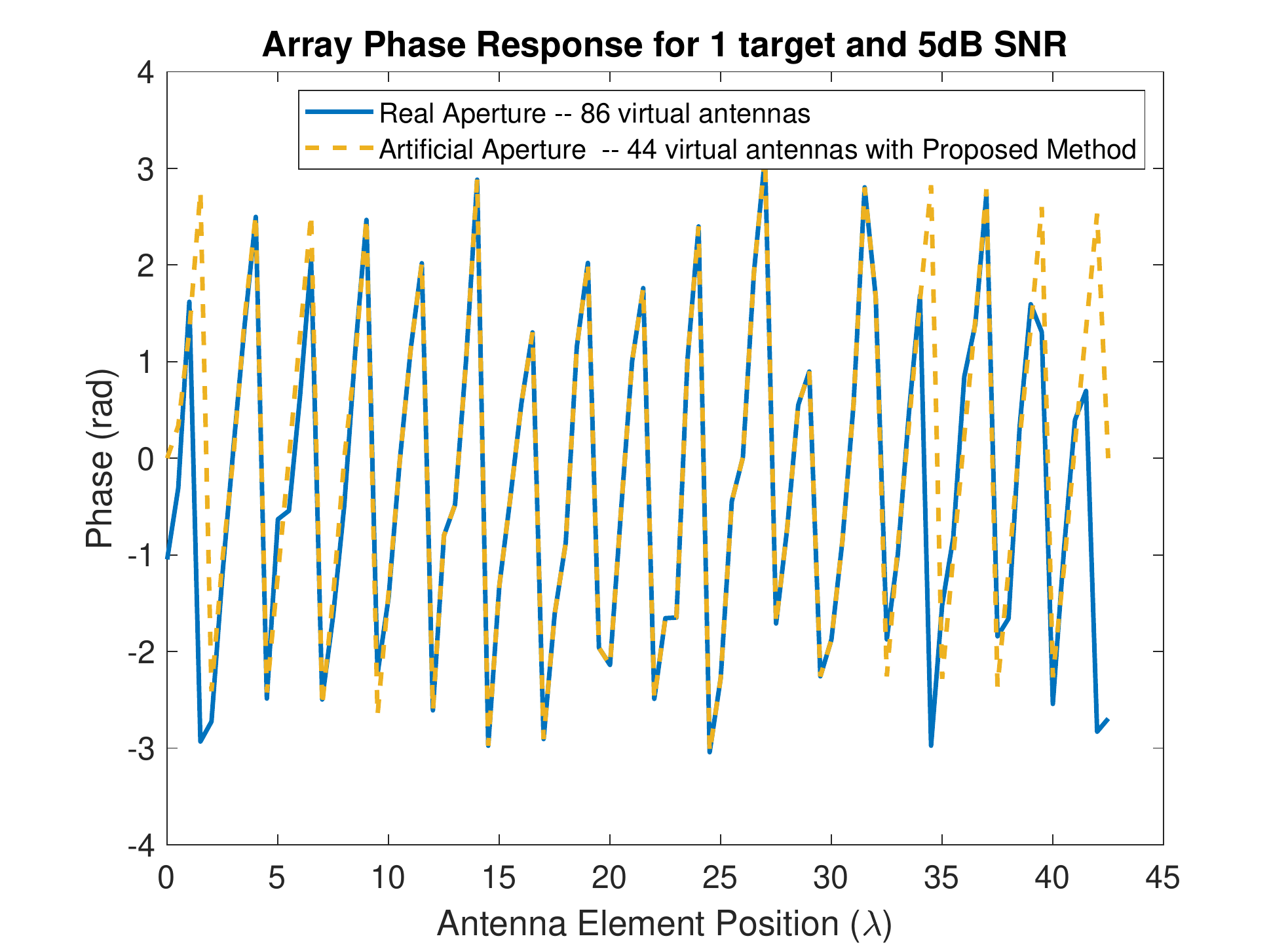}
\caption{Signal phase for the large aperture radar (blue) and for the enhanced version of the small aperture radar using the proposed method (yellow).}
\label{Phase}
\end{figure}

Since the angle information is encoded in the signal phase, the accurate extrapolation of the phase is a requirement for the method to work. As an example of this phase reconstruction, Fig.~\ref{Phase} shows the signal phase for a test scene with only one point target and 5 dB SNR, for the array with the large aperture and the array with a small aperture processed with the proposed method. It can be seen how the phase is well extrapolated, which will lead to a higher angular resolution estimation. However, it can also be seen that the phase is not perfectly reconstructed, which could lead to higher side-lobe levels or false alarms in the worst case. For this reason, a statistical analysis of the 50.000 test cases has been done, addressing the accuracy in the angle estimation, the minimum angular separation, and the probability of false alarms  ($P_{fa}$).

\begin{figure}
\includegraphics[width=\linewidth]{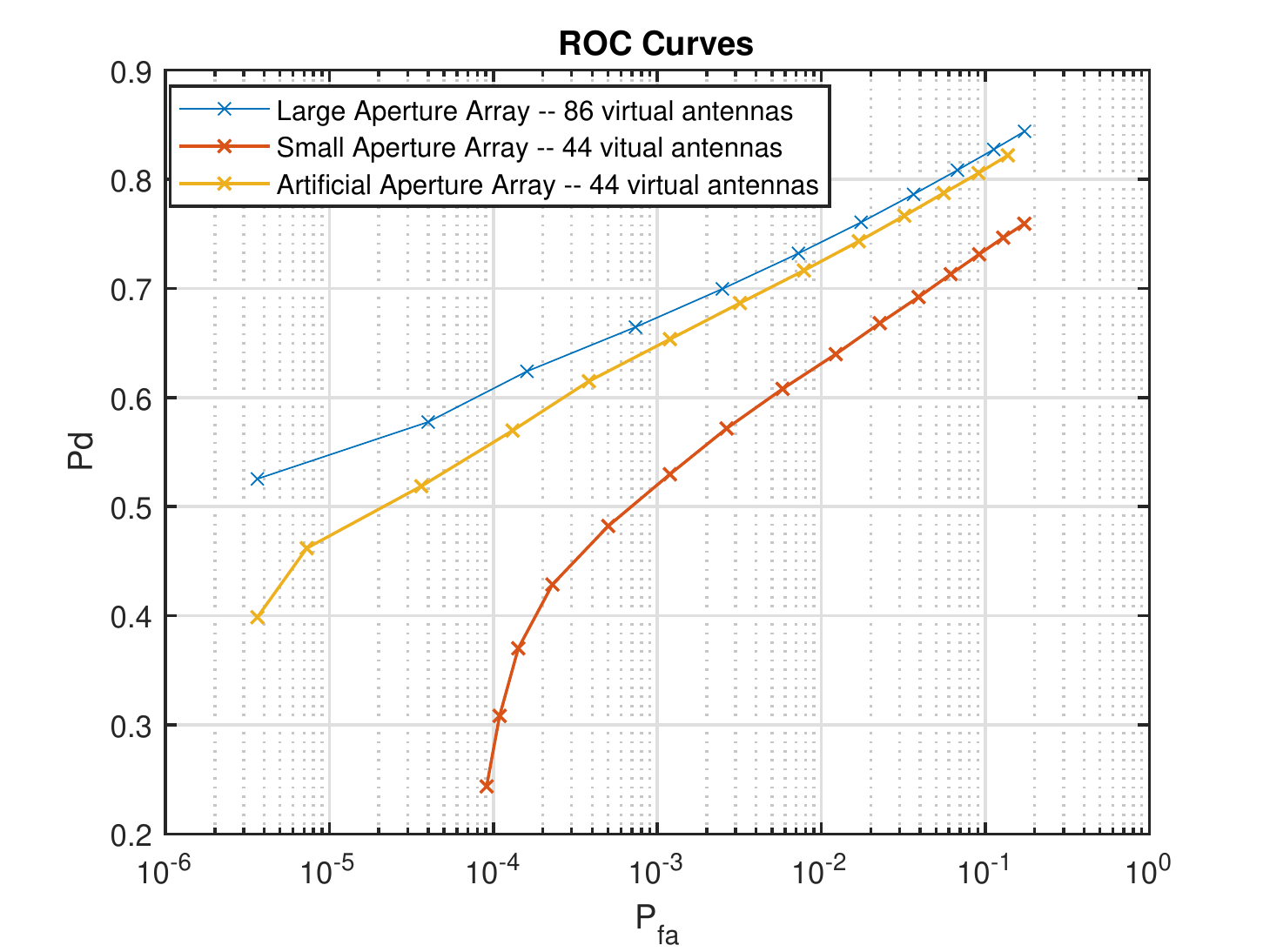}
\caption{ROC curves using a fixed threshold detector after a Fourier beamformer. In blue, the detection capabilities of the large aperture array. In red, the detection capabilities of the small aperture array. As expected, the performance of the small array is lower, both because of the resolution degradation and the decrease in the SNR due to coherent integration. In yellow, the enhanced version with the proposed method using only 44 antennas.}
\label{ROC}
\end{figure}

First, a ROC curve has been computed using a fixed threshold detector, where the threshold is varied by reducing its value from the maximum. For simplicity, a conventional beamformer, also know as Fourier beamformer, has been used for DoA estimation. Since the positions of the targets are generated randomly, in many cases the angular separation between them will be too small to trigger two detections even for the large aperture array. Moreover, in some scenes the SNR is set to -5 dB, and weak targets will be present. For these two reasons the probability of detection ($P_d$) never reaches 1. Fig.~\ref{ROC} shows the ROC curves computed for the 50.000 test cases. Two observations can be made from this plot. First, the $P_{fa}$ is not increased in the artificial aperture array, and therefore, no significant undesirable effects are introduced by the proposed method. Second, the $P_d$ is higher in the artificial aperture array than in the small array, even though they use the same number of physical antennas. This effect can be explained that due to the increase in angular resolution, more targets are correctly detected in situations where the targets are closely spaced.

\begin{figure*}[!t]
\centering
\subfloat[]{\includegraphics[width=3in]{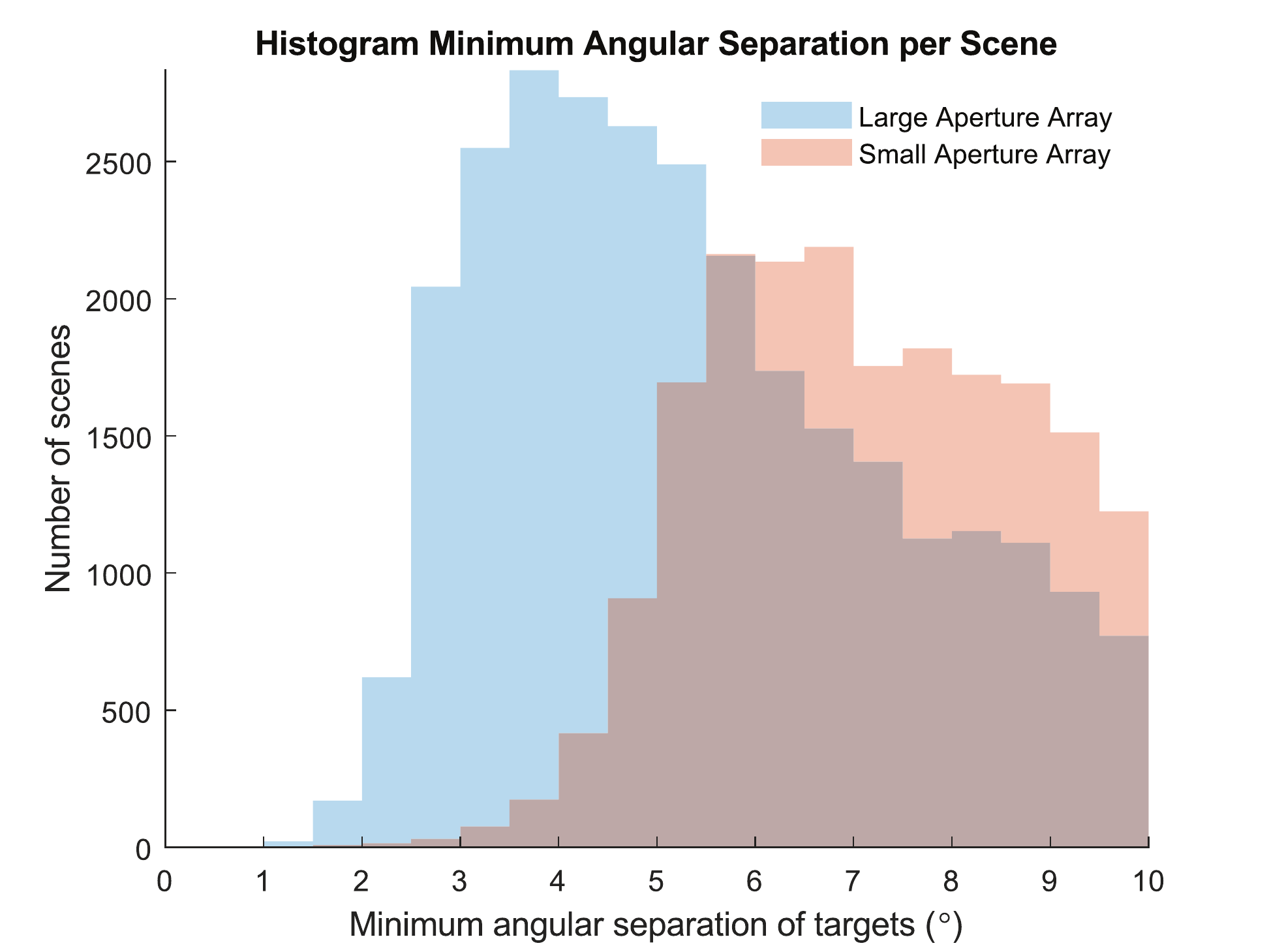}%
\label{HistTrim}}
\hfil
\subfloat[]{\includegraphics[width=3in]{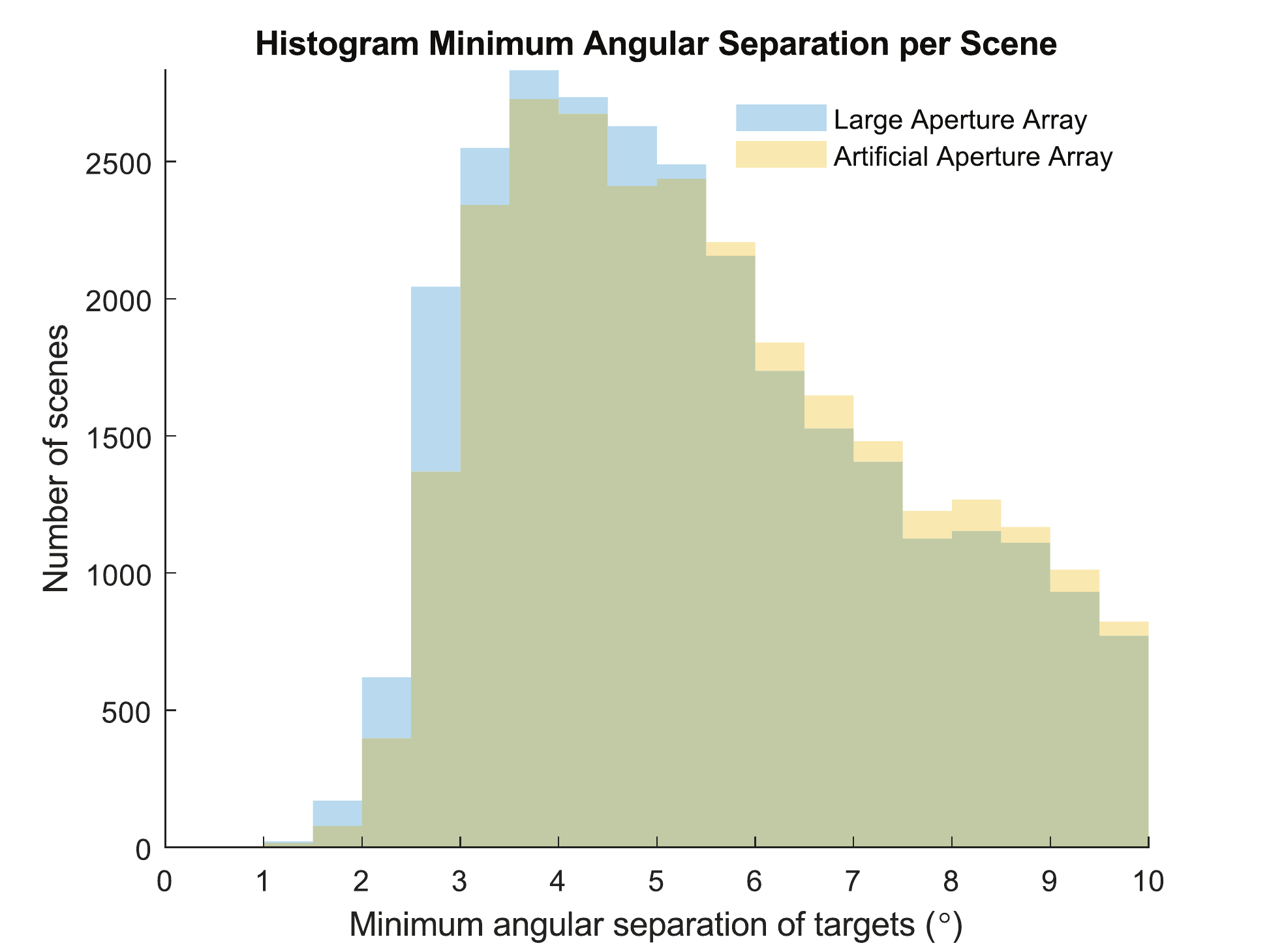}%
\label{HistRe}}
\caption{Histograms of the minimum detected angular separation per scene. (a) compares the large and small aperture array, and as expected, the small aperture array cannot resolve very close-spaced targets. (b) compares the large aperture array and the proposed method applied to the small array. It can be seen how the two distributions are more similar, meaning that the proposed method enhances the angular resolution.}
\label{Histograms}
\end{figure*}

However, the increase of the $P_d$ for a fixed $P_{fa}$ is also induced by the increase of the SNR when processing more coherent antennas, as the target's power will add coherently while the noise will not. For this reason, it is important to analyze the minimum angular separation between detections independently. For each scene, the minimum angular separation has been computed as:

\begin{equation}
    min(|\hat{\theta_i} - \hat{\theta_j}|) \quad \forall i,j \in [1, N_{targets}] \ | \;  i \neq j,
    \label{minDistance}
\end{equation}

where $\hat{\theta_i}$ is the estimated DoA of the target $i$ and $N_{targets}$ is the number of targets in the scene. The histograms of the minimum angular separation of the detected targets are presented in Fig.~\ref{Histograms}. In Fig.~\ref{HistTrim}, it can be seen how the large aperture array is able to resolve targets that are more closely spaced than the small aperture array. This is the expected behavior, since it is known from \eqref{resolution} that the (boresight) angular resolution of a 86 ULA and a 44 ULA are 1.33$^\circ$ and 2.6$^\circ$, respectively. On the other hand, Fig.~\ref{HistRe} shows the minimum angular separation when applying the proposed method to the small aperture array. It can be seen how the two distributions are more similar, indicating that the proposed method can improve the performances of the small physical array making it more similar to those of the physically larger array. Nevertheless, the distribution for the artificial aperture array is more skewed to a higher angular separation than the original large array, but still performs better than the small aperture array in most cases.

Finally, it is important to analyze if the method induces a degradation in the accuracy of the DoA estimation. In this case, two DoA estimation methods have been evaluated, the same Fourier beamformer used for the previous results in this section, and the single snapshot MUSIC proposed in \cite{LIAO201633}. The Mean Squared Error (MSE) of the angular estimation for each method has been computed, and the results have been averaged over the test scenes. Moreover, the Cramer-Rao Bound (CRB) has been included for comparison, calculated as \cite{vantrees2002}:

\begin{equation}
    CRB = \frac{\sigma^2_n}{2}
    \biggl| Re \Bigl[ S^H D^H(I-A(A^HA)^{-1}A^H)DS \Bigr]\biggr| ^ {-1}
    \label{CRb}
\end{equation}
where:
\begin{itemize}

\item $A = [v(\theta_1), ... , v(\theta_p)]$ is the array steering matrix, formed by $p$ steeting vectors. 
\item $D = [d(\theta_1), ... , d(\theta_p)]$ contains the first derivative of steering vectors $d(\theta) = \frac{\delta v(\theta_1)}{\delta\theta}$ 
\item $S = diag(s_1, ..., s_p)$ formed by the $p$ targets complex response.
\item $\sigma^2_n$ is the noise variance.
\end{itemize}

\begin{figure}
\includegraphics[width=\linewidth]{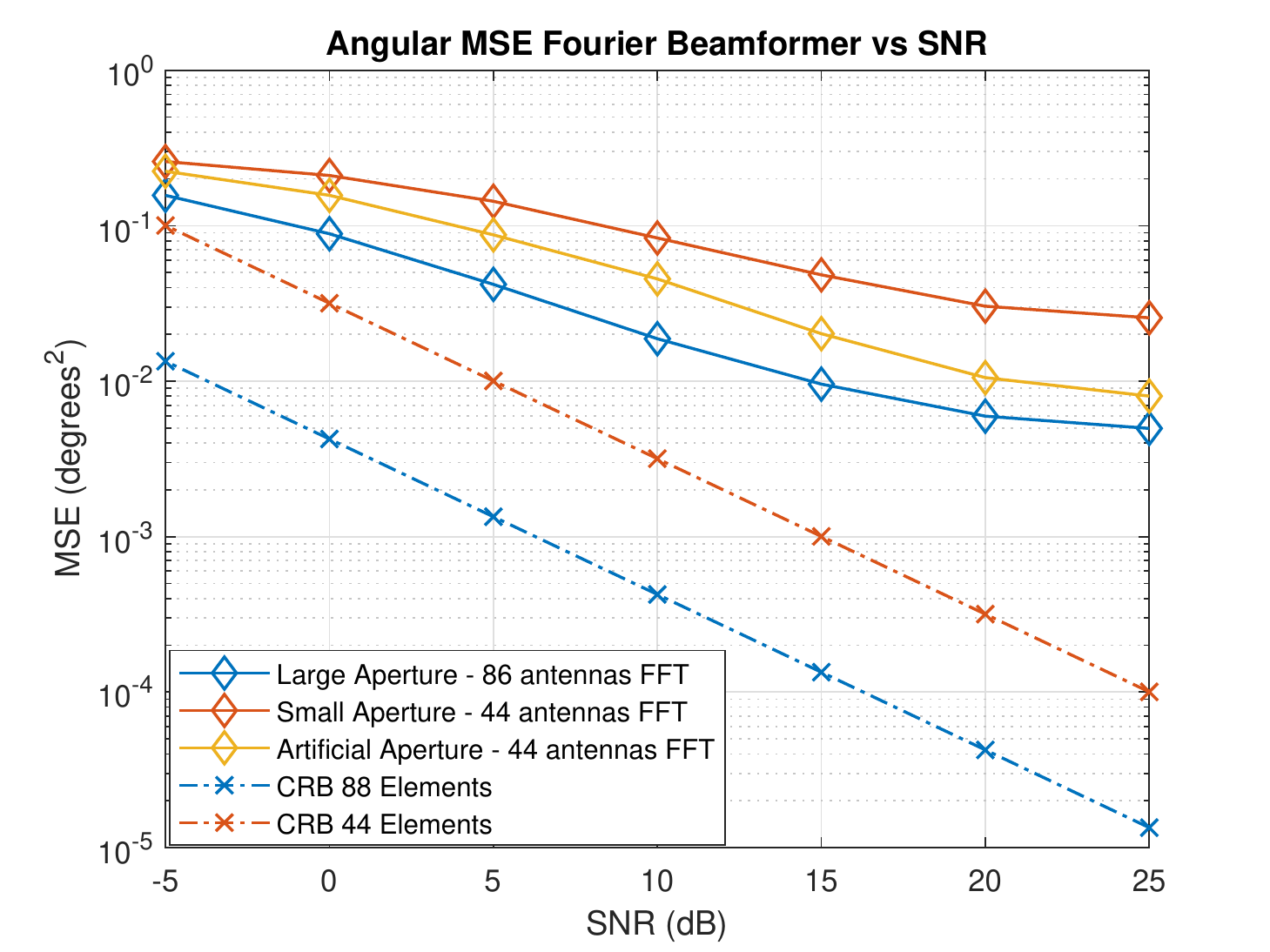}
\caption{MSE of the Fourier beamformer. In blue are shown the results for the 86 antenna array with its corresponding CRB. In red are shown the results for the 44 antenna array and CRB. In yellow is shown the performance when applying the proposed method to the 44 antenna array before the beamforming.}
\label{MSE_CRB_FFT}
\end{figure}

\begin{figure}
\includegraphics[width=\linewidth]{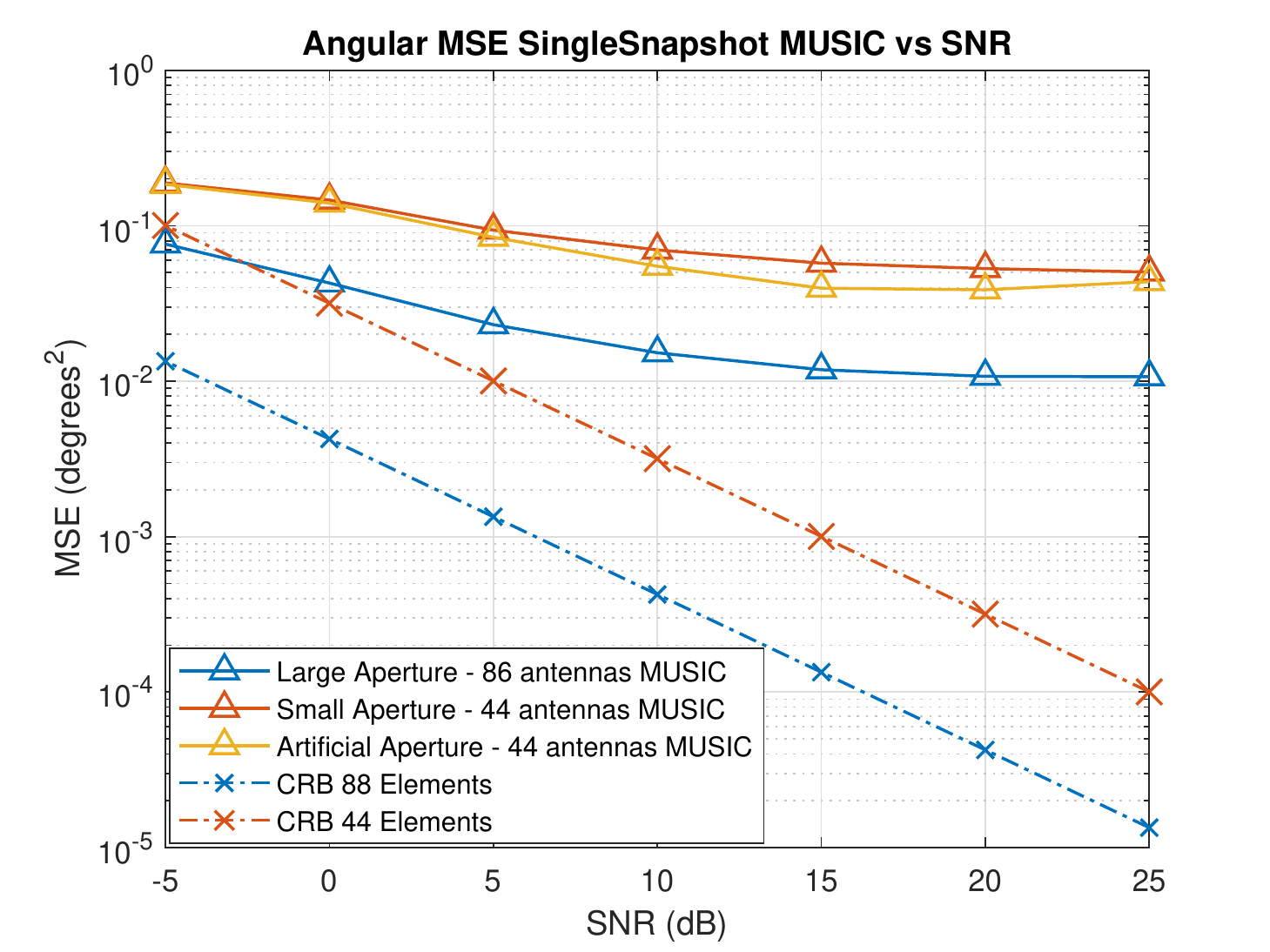}
\caption{MSE of the single snapshot MUSIC algorithm. In blue are shown the results for the 86 antenna array with its corresponding CRB. In red are shown the results for the 44 antenna array and CRB. In yellow is shown the performance when applying the proposed method to the 44 antenna array before the MUSIC algorithm.}
\label{MSE_CRB_MUSIC}
\end{figure}

The results are shown in Fig.~\ref{MSE_CRB_FFT} for the Fourier beamformer and in Fig.~\ref{MSE_CRB_MUSIC} for the single snapshot MUSIC. In both cases, it can be seen how the artificial aperture generated with the proposed method has an improvement in the MSE with respect to the small aperture array. However, the improvement in the single snapshot MUSIC is lower. This is due to the fact that the MSE is only computed in those cases where the targets are resolved correctly. Since the single snapshot MUSIC requires the exact number of targets in the scene as input, it resolves more challenging cases than the Fourier beamformer (i.e., cases that in a real unknown scene would not be resolved). Thus, cases where the DoA angles to be estimated are very close are included for the single snapshot MUSIC evaluation, degrading the method's performance.

\section{Evaluation on Experimental Data}
To evaluate the performance of the presented framework, experimental data from different scenarios have been collected with a commercial FMCW MIMO radar (Texas Instrument MMWCAS-RF), working at 79GHz and capable of synthesizing 86 virtual antennas by cascading four conventional radar chips \cite{Tidep}. The details of the parameters used to collect the data can be seen in Table II. The data collection was designed to be heterogeneous in the type of targets and the relative position between them and with respect to the radar. Also, different scenarios with different background clutter have been included.

\begin{table}[!t]

\caption{Radar Parameters for Experimental Validation\label{RadarParameters}}
\centering
\begin{tabular}{|c|c|}
\hline
Center Frequency & 78.58 GHz\\
\hline
Effective Bandwidth & 320 MHz \\
\hline
Chirp slope &  5 MHz / \textrm{$\mu$}s \\
\hline
Samples per chirp & 256 \\
\hline
Chirp period & 80 \textrm{$\mu$}s \\
\hline
Chirps per frame & 128\\
\hline
Transmitted power & +13 dBm \\
\hline
Number of effective azimuthal antennas & 86 \\
\hline
\end{tabular}
\end{table}

\begin{figure}
\includegraphics[width=\linewidth]{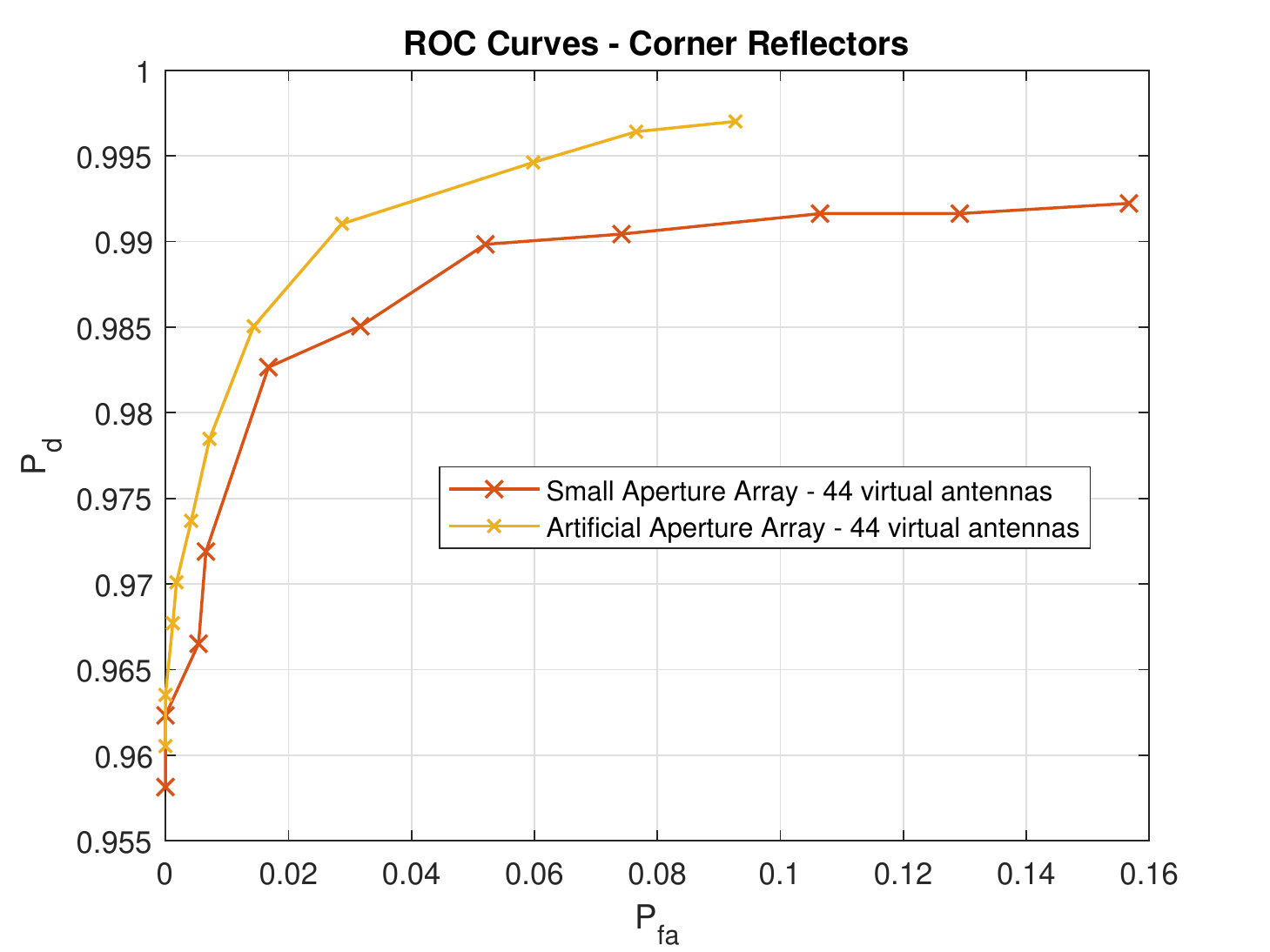}
\caption{ROC curves for 25 scenes with a different number of corner reflectors using the estimation computed with the large aperture array as ground truth. In red, the performance when only 44 virtual antennas are used. In yellow, the performance when the proposed method is applied before the angle estimation.}
\label{ROC_Exp}
\end{figure}

First, an analysis of the performance using corner reflectors has been done. To this end, 25 scenes have been recorded in two different scenarios, placing corner reflectors with different RCS at the same range and different angles. The detections obtained with the full array have been used as ground truth in this case, to avoid errors induced by the lack of equipment for measuring the exact positioning of the targets. Similar to the simulated scenes in the previous section, the ROC curve averaging the results for the 25 recordings has been computed. Fig.~\ref{ROC_Exp} shows the result when performing the DoA estimation using only 44 virtual antennas and the result when the trained NN is used beforehand to artificially augment the resulting aperture. The same data has been used for generating both curves, which consist of 11 points each. It is important to mention that the NN has only been trained with simulated data to push its generalization capabilities to unseen, in this case experimental, data. As it can be seen, the artificial aperture, enhanced with the NN, can have better detection probability while keeping the probability of false alarm low. This is because the small aperture is merging targets together due to the limited angular resolution; on the other hand, the proposed method can help separate such targets.

\begin{figure}
\includegraphics[width=\linewidth]{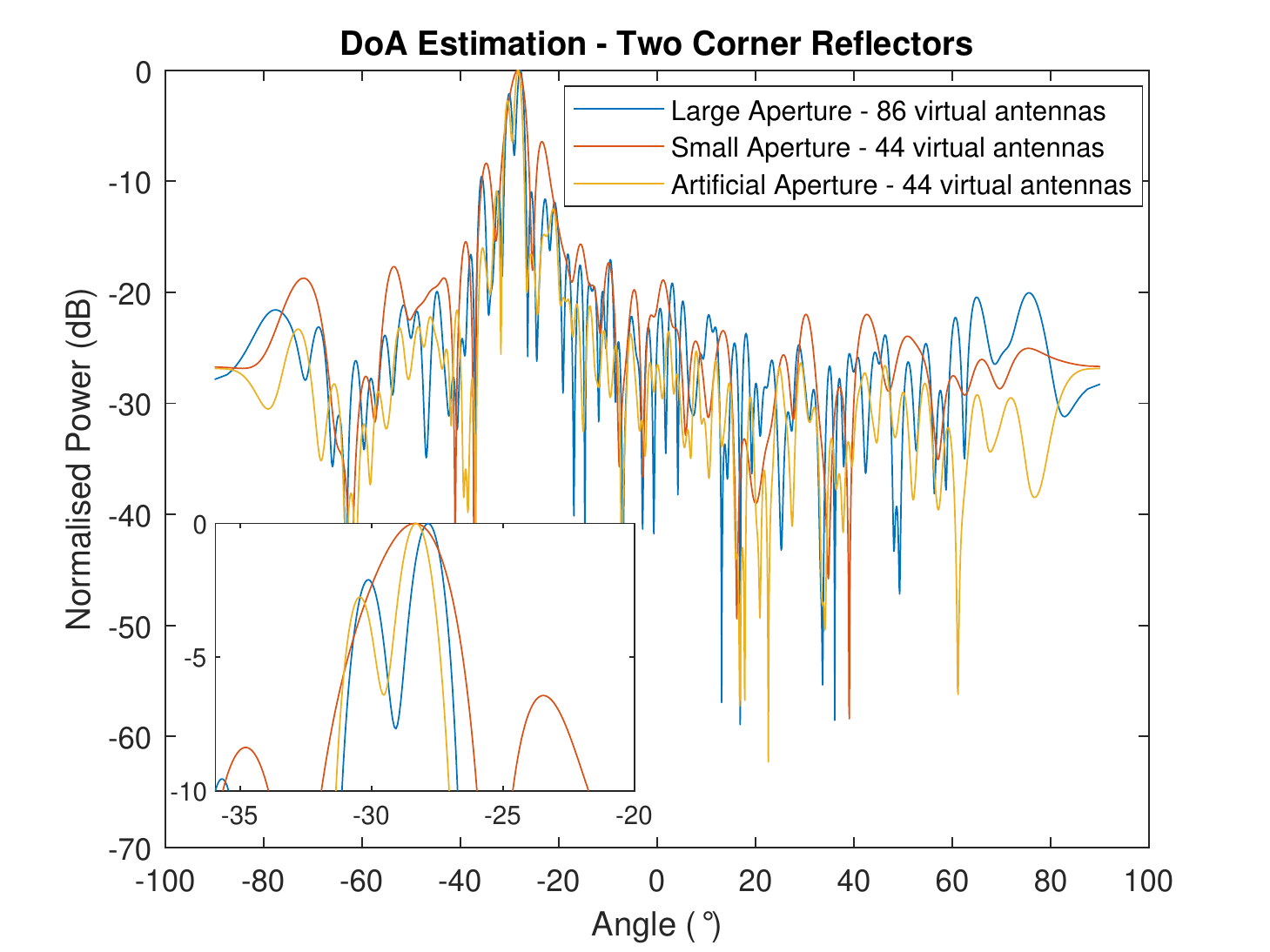}
\caption{Angle estimation of a scene with two corner reflectors with a small angular separation. In blue, the estimation with the large aperture array; in red, the estimation with the small aperture array; in yellow, the estimation for the artificial aperture enhanced with the proposed method.}
\label{DoA2Corner}
\end{figure}

Moreover, an example of the angular domain estimation is shown in Fig.~\ref{DoA2Corner}, where two corner reflectors are placed at -30 degrees in the same range. As it can be seen, when the angular domain is estimated using the whole array ('full aperture'), two main peaks are distinguishable. However, when only 44 virtual elements are used emulating a smaller aperture array, these two peaks are merged due to poor angular resolution. On the other hand, if the proposed method is applied before the angle estimation, the two peaks can be clearly seen again ('artificial aperture').
It is also possible to observe that the peaks in the artificial aperture estimation are slightly shifted from the full version. The average estimation accuracy using the full aperture estimation as ground truth has been computed for the small and artificial apertures. The mean MSE over the 25 scenarios is 0.0268 for the small aperture and 0.0075 for the artificial aperture. These numbers confirm the simulation results, where the proposed method does not degrade the accuracy performance but improves it.

\begin{figure}
\includegraphics[width=\linewidth]{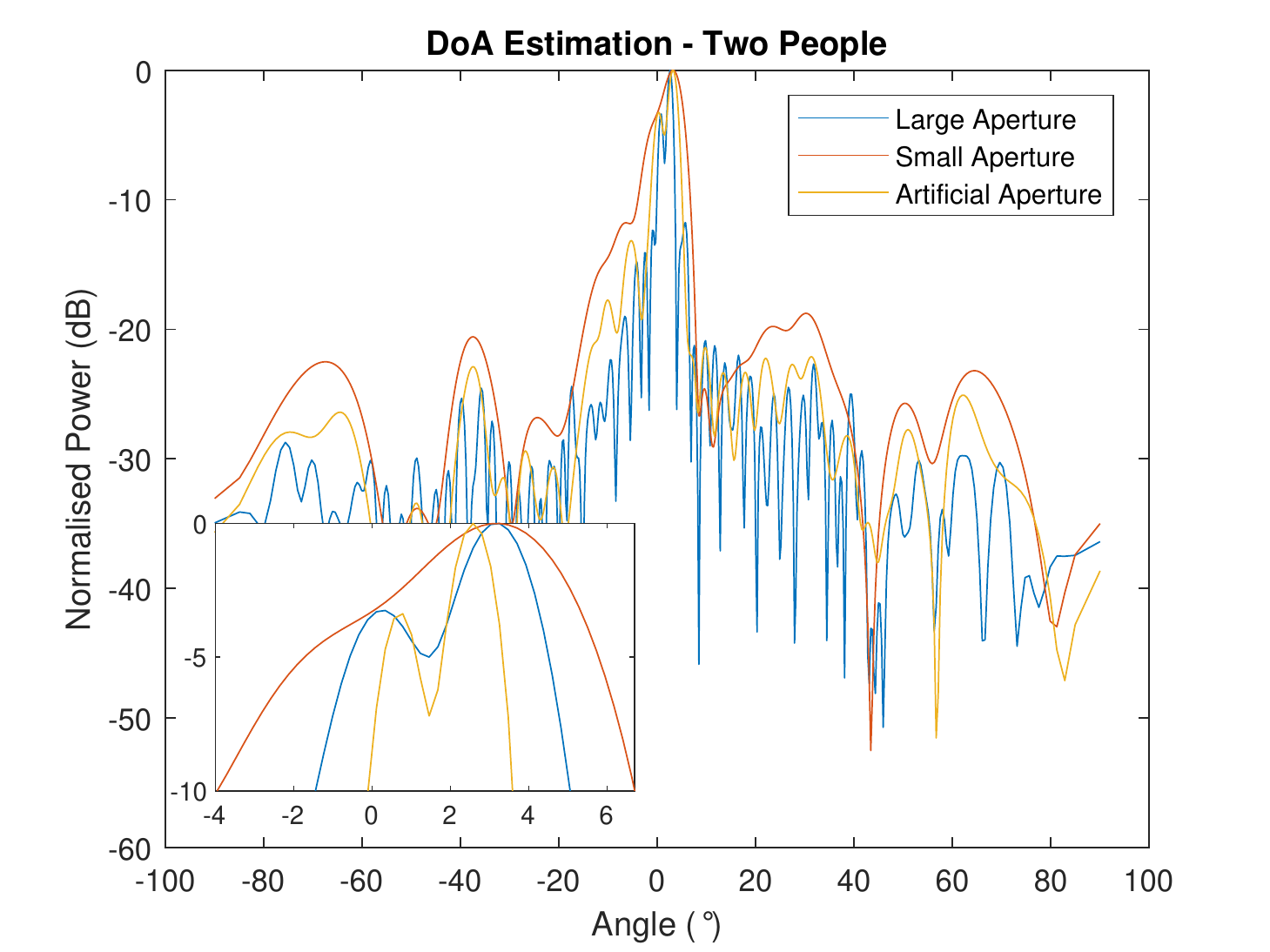}
\caption{DoA estimation of a scene with two people with a small angular separation. In blue, the estimation with the original large aperture array; in red, the estimation with the small aperture array; in yellow, the estimation with the artificial aperture enhanced via the proposed method.}
\label{DoAPeople}
\end{figure}

\begin{figure*}[!t]
\centering
\includegraphics[width=\textwidth]{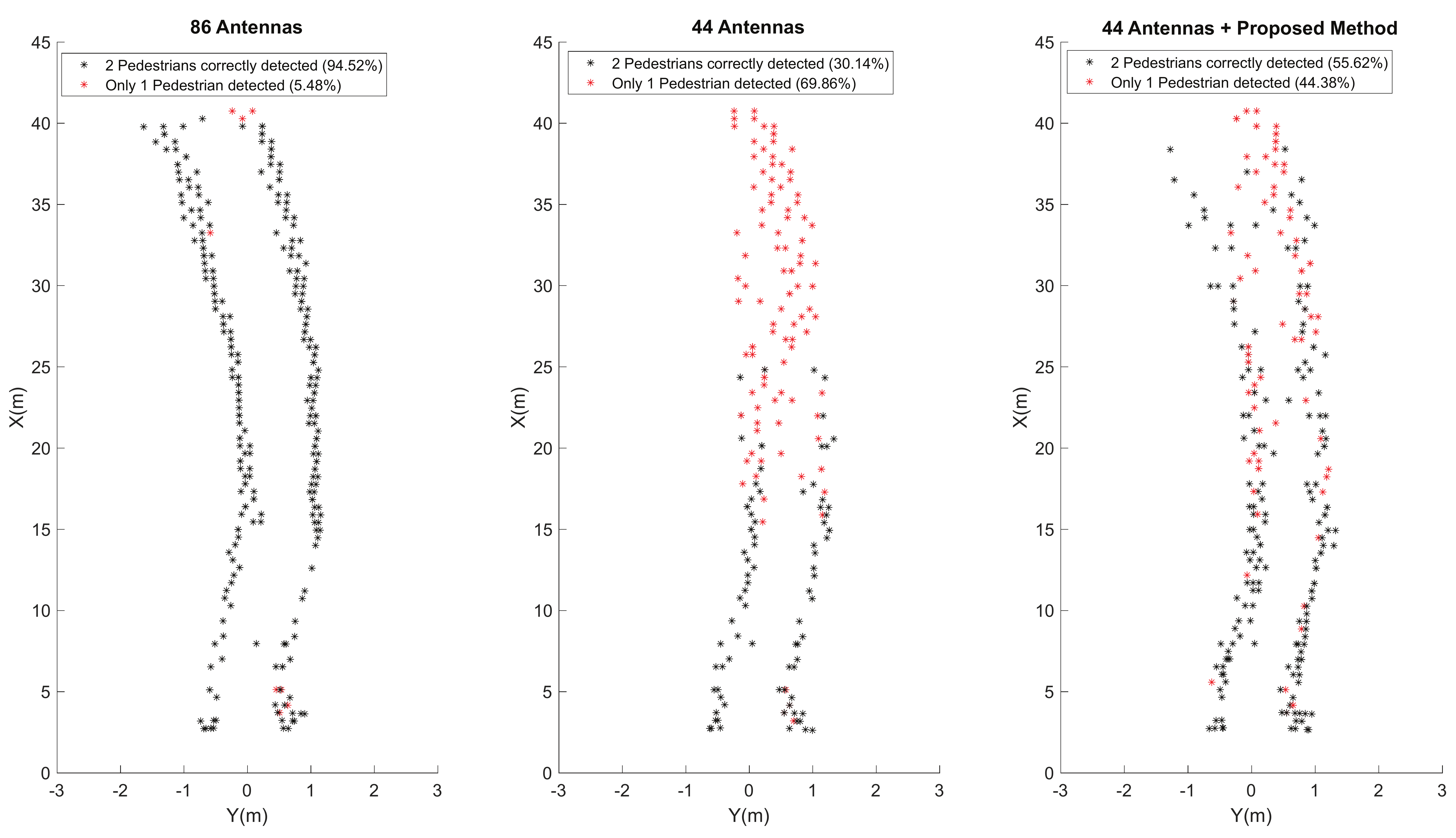}
\hfil
\caption{Detections triggered in a scene where two people walk toward the radar. If two detections are triggered, meaning that both people have been detected, the detections are plotted in black. On the other hand, if only one peak is detected, the detection is plotted in red. More than 94\% of the snapshots are correctly resolved with the full aperture array. With the small aperture array, only 30\% are correctly detected, while this number is boosted to 55\% with the proposed method.}
\label{Pedestrians}
\end{figure*}

Finally, the proposed method has been tested with more realistic automotive targets. To this end, a new set of measurements have been collected where two people are present in the scene. Fig.~\ref{DoAPeople} shows one angle estimation example where the small aperture radar cannot generate two peaks corresponding to the two pedestrians. This is a crucial result since the radar will report a single large object instead of two independent targets, which may lead to critical errors in the subsequent decision-making system. To illustrate this phenomenon better, a dynamic experiment has been carried out,  where two pedestrians have walked towards the radar at the same speed and range, maintaining a constant separation of approximately 1 meter between them in the angular direction. After applying standard radar processing and filtering out the static clutter, the estimation of the angular domain has been done with an FFT for the large, the small, and the artificial aperture. In each snapshot a simple peak detector has been applied, and the results have been plotted in Fig.~\ref{Pedestrians}. If two detections are triggered in the snapshot (i.e., the two pedestrians have been detected), the detections are plotted in black and are considered as correct detections. On the other hand, if only one peak is found, the two pedestrians have been merged due to poor angular resolution; thus, the detection is plotted in red and counted as an incorrect detection. As seen in the left-hand side of Fig.~\ref{Pedestrians}, when using the 86 virtual antennas, the trajectory of both pedestrians can be clearly followed since in more than 94\% of the snapshots both people are detected. However, when the angular domain is estimated with only 44 virtual antennas, both pedestrians started to merge into a single detection after approximately 15 meters, and became completely indistinguishable after 25 meters, as seen in the middle Fig.~\ref{Pedestrians}. In this case, only in 30\% of the snapshots can both people be detected. Finally, if the method presented in this paper is applied before estimating the angular domain, the percentage of correct snapshots is boosted to 55.62\%. This effect can be seen in the right-hand side of Fig.~\ref{Pedestrians}, where both trajectories are again discernible. These results show a clear benefit of applying the proposed method since the enhanced radar system can distinguish the two targets at a greater distance. Thus, the decision-making system of the autonomous vehicle will have more time to react.

\section{Conclusions}
This paper presents a novel framework to enhance the angular resolution in MIMO radars without increasing the number of physical antennas. A Neural Network with a self-supervised learning scheme is used to extrapolate the antenna array's response, increasing its aperture artificially. Data from a high angular resolution radar (i.e., a radar with a large aperture) is used to train an LSTM Network that is later used to increase the angular resolution of a low-resolution radar (i.e., a radar with a small aperture). One million simulated scenes have been used to train the system, validating its performance in terms of detection capabilities and estimation accuracy. Two angular estimators have been used, a simple Fourier beamformer and the single snapshot MUSIC, and their performances have been compared with and without applying the proposed method. This demonstrated that the proposed method can enhance the angular resolution of MIMO radars without introducing false alarms or degradation in the estimation accuracy.

Moreover, results in experimental data show that this method can help in automotive scenarios to resolve closely spaced targets such as pedestrians walking closely together, providing accurate information earlier to the decision-making system. To the best of the author's knowledge, this is the first time that Neural Networks have been used to increase the angular resolution of MIMO radars. Finally, it is important to mention that the presented work aims to propose a framework in which different Neural Network architectures could be used, but the exploration of their different models and detailed architectures is out of the scope of this paper.

\section*{Acknowledgments}
The authors are grateful for the constructive comments of the anonymous reviewers and editors, which improved the quality of this work and also to all the members of the MS3 group that were involved in the data acquisition campaigns.

\bibliographystyle{IEEEtran}
\bibliography{refs}

\newpage

\begin{IEEEbiography}[{\includegraphics[width=1in,height=1.25in,clip,keepaspectratio]{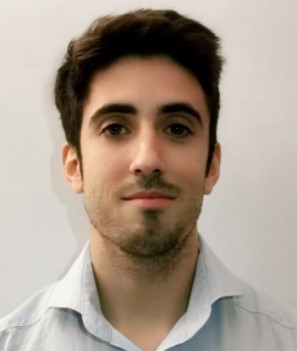}}]{Ignacio Roldan}
received his B.Sc. and M.Sc. in Telecommunication Engineering at the Universidad Politécnica de Madrid, Spain in 2014 and 2016. In 2018 he complemented his education with an M.Sc. in Signal Processing and Machine Learning at the same university. He has worked more than 5 years at Advanced Radar Technologies, a Spanish tech company focused on the design and manufacture of radar systems. During this period, he has been involved in several international projects developing state-of-the-art signal processing techniques for radars. In his last stage, he was focused on applying Machine Learning techniques to UAV detection and classification. In Sep 2020, he joined the Microwave Sensing, Signals, and Systems group at Delft University of Technology, where he is working towards the PhD degree.
\end{IEEEbiography}

\begin{IEEEbiography}[{\includegraphics[width=1in,height=1.25in,clip,keepaspectratio]{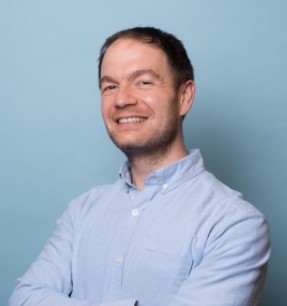}}]{Francesco Fioranelli}
(M'15–SM'19) received the Ph.D. degree with Durham University, Durham, UK, in 2014. He is currently an Associate Professor at TU Delft, The Netherlands, and was an Assistant Professor with the University of Glasgow (2016–2019), and a Research Associate at University College London (2014–2016). His research interests include the development of radar systems and automatic classification for human signatures analysis in healthcare and security, drones and UAVs detection and classification, automotive radar, wind farm, and sea clutter. He has authored over 140 publications between book chapters, journal and conference papers, edited the books on “Micro-Doppler Radar and Its Applications” and "Radar Countermeasures for Unmanned Aerial Vehicles" published by IET-Scitech in 2020, and received three best paper awards.
\end{IEEEbiography}

\begin{IEEEbiography}[{\includegraphics[width=1in,height=1.25in,clip,keepaspectratio]{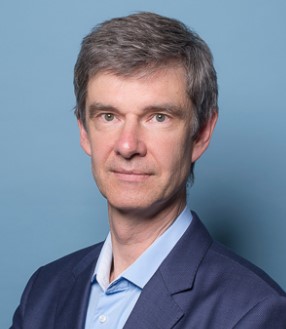}}]{Alexander Yarovoy}
(FIEEE'15) graduated from the Kharkov State University, Ukraine, in 1984 with the Diploma with honor in radiophysics and electronics. He received the Candidate Phys. \& Math. Sci. and Doctor Phys. \& Math. Sci. degrees in radiophysics in 1987 and 1994, respectively.
In 1987 he joined the Department of Radiophysics at the Kharkov State University as a Researcher and became a Full Professor there in 1997. From September 1994 through 1996 he was with Technical University of Ilmenau, Germany as a Visiting Researcher. Since 1999 he is with the Delft University of Technology, the Netherlands. Since 2009 he leads there a chair of Microwave Sensing, Systems and Signals. His main research interests are in high-resolution radar, microwave imaging and applied electromagnetics (in particular, UWB antennas). He has authored and co-authored more than 450 scientific or technical papers, six patents and fourteen book chapters. He is the recipient of the European Microwave Week Radar Award for the paper that best advances the state-of-the-art in radar technology in 2001 (together with L.P. Ligthart and P. van Genderen) and in 2012 (together with T. Savelyev). In 2010 together with D. Caratelli Prof. Yarovoy got the best paper award of the Applied Computational Electromagnetic Society (ACES).
Prof. Yarovoy served as the General TPC chair of the 2020 European Microwave Week (EuMW'20), as the Chair and TPC chair of the 5th European Radar Conference (EuRAD'08), as well as the Secretary of the 1st European Radar Conference (EuRAD'04). He served also as the co-chair and TPC chair of the Xth International Conference on GPR (GPR2004). He served as an Associated Editor of the International Journal of Microwave and Wireless Technologies from 2011 till 2018 and as a Guest Editor of five special issues of the IEEE Transactions and other journals. In the period 2008-2017 Prof. Yarovoy served as Director of the European Microwave Association (EuMA)
\end{IEEEbiography}

\vfill

\end{document}